\documentclass[aps,twocolumn,showpacs,floatfix,amssymb]{revtex4}
\usepackage{graphicx}
\begin{document}

\title{On the  Coulomb interaction in\\
superconducting pairing in cuprates }
\author{Nikolay M. Plakida }
 \affiliation{Joint Institute for Nuclear Research,
 141980 Dubna, Russia}

\date{\today}

\begin{abstract}

We discuss a reduction of   superconducting transition temperature by the intersite Coulomb repulsion for the  $d$-wave pairing  in cuprates.   We compare the results found for  the spin-fermion model and the
extended Hubbard model. We argue that in both the models the  $d$-wave superconducting
transition temperature is reduced by the Coulomb repulsion of holes in different unit
cells. We also show that in the strong correlation limit the  $s$-wave superconductivity
cannot occur due to the kinematic restriction  of no double occupancy in the Hubbard
subbands.
\end{abstract}


 \maketitle

It is commonly believed that the Coulomb interaction  is detrimental to
superconductivity. In particular, for low-temperature superconductors with the $s$-wave
pairing  mediated by electron-phonon coupling the retardation effect renormalizes the
Coulomb interaction which makes it feasible to obtain a finite superconducting
$T_c$~\cite{Bogoliubov58,Morel62}. For  electronic pairing mechanisms,  the retardation
effect is ineffective and the Coulomb interaction suppresses the $s$-wave pairing. Only
superconducting pairing  with higher orbital momenta,  $p, d, f, \ldots $, can occur in
the Fermi-liquid as originally was proposed by Kohn and Luttinger~\cite{Kohn65} (for a
review see \cite{KaganM16}).

The cuprate superconductors are the Mott-Hubbard (more accurately, charge-transfer) doped
insulators caused by the large Coulomb interaction $U_d$ on copper sites.  In this case
the interaction  $U_d$ should be taken into account rigorously  in considering the
electronic structure of cuprates.  The most frequently  used  is the three-band $p$--$d$
model for  Cu$\, 3\, d(x^{2}-y^{2})$-states and O$2p_{\sigma} (x,y)$-states in the
CuO$_2$ plane~\cite{Emery87,Varma87}.  To obtain a tractable model for description of
low-energy electronic excitations the  $p$--$d$ model can be reduced to  simpler  models.

In particular,   in the spin-fermion model (SFM) the high-energy excitations on copper
sites are excluded, which results in a conduction band  for oxygen holes on the O$2p_{\sigma}
(x,y)$ orbitals interacting with localized copper spins $S= 1/2$ in the CuO$_2$ plane
(see, e.g.,~\cite{Barabanov88}). In Ref.~\cite{Valkov16}, the SFM was used to consider the
$d$-wave   superconducting pairing for spin-polarons. It was found that the Coulomb
interaction $V_{pp}$ between holes on the nearest neighbor oxygen sites gives no
contribution to the $d$-wave pairing by  symmetry reason.  This was considered as a proof
of stability of the $d$-wave pairing towards  the intersite Coulomb repulsion. However,
the authors have neglected the Coulomb interaction between  holes in  different unit
cells,  which conventionally reduces  the superconducting $T_c$.

Another approach is   based on the cell-cluster perturbation theory (see
\cite{Hayn93,Belinicher94,Feiner96,Yushankhai97,Korshunov05}).  In the theory,  the
spectrum of  electronic excitations in the unit cell  CuO$_4$ is rigorously calculated by
an  exact diagonalization of the copper and oxygen energy states taking into account all
relevant  Coulomb interactions, $U_d,\, U_p,\, V_{pd}, \, V_{pp}$, and hybridizations,
$t_{pd}, \, t_{pp}$. Considering the lowest energy states close to the Fermi level, the
singly occupied $d (x^2 - y^2)$ states  and doubly occupied singlet $p$-$d$ hole states,
the extended Hubbard model (EHM) can be formulated  for two  Hubbard subbands with the
hopping parameter between different unit cells   $ t \sim 0.3 t_{pd}$ and the intersite
Coulomb repulsion $V \sim 0.5\, t $~\cite{Feiner96}.

In the limit of strong correlations the projected (Hubbard) electronic operators should
be used~\cite{Hubbard65}. They have nonfermionic commutation relations, as e.g., the commutation relation for the Hubbard operators in the singly occupied subband, $ X_i^{0 \sigma } = a_{i\sigma}(1-
N_{i\bar\sigma})$ where $N_{i \sigma} = a^\dag_{i\sigma} a_{i\sigma}, \; \sigma = \pm 1,\, \bar\sigma = -
\sigma $, reads
\begin{equation}
X_i^{0 \sigma }   X_j^{\sigma0} +  X_j^{\sigma0}X_i^{0 \sigma }  = \delta_{ij}(1 - N_{i
\sigma} /2 + \sigma S_i^z).
   \label{1a}
\end{equation}
This results in the kinematic interaction for electrons,  which is determined by electron
scattering on charge (number $N_{i \sigma}$) and spin $S_i^\alpha$ fluctuations with the
coupling of an order of the hopping parameter $\, t$.

In Refs.~\cite{Plakida13,Plakida14}, the EHM   was studied within the strong coupling
superconducting theory. It was shown that the spin-fluctuation  pairing induced by the
kinematical interaction in the second order of $\, t \,$ results in the $d$-wave
superconductivity with high-$T_c$. In Ref.~\cite{Plakida13}, it was found  that the
intersite Coulomb repulsion $V \sim 0.5\, t $ in cuprates is not strong enough to
suppress  the $d$-wave superconductivity.  To prove this pairing mechanism, in
Ref.~\cite{Plakida14} we consider a much stronger than in cuprates Coulomb interaction
$V$. Only for $V$  larger than the coupling constant for the spin-fluctuation pairing, $V  \gtrsim 4\,t$,  the $d$-wave pairing can be fully suppressed.

Now we comment on  the $s$-wave superconducting pairing in  cuprates.    The $s$-wave
pairing induced by the kinematical interaction in the limit of strong correlations was
originally proposed  in Refs.~\cite{Zaitsev87}.  In Refs.~\cite{Zaitsev04, Valkov11},  the superconducting pairing in the two-dimensional Hubbard model  was studied. It
was found that this pairing is robust in respect to the  intersite  Coulomb interaction
$V$. However, in these papers only the first order of the kinematical interaction $\propto t $ was
considered, which results in the  $s$-wave  pairing with the symmetric superconducting gap
 $ \,\Delta_{\sigma}(q_x, q_y) =  \Delta_{\sigma}(q_y,
q_x)\,$. However, in this case the well known constraint of ``no double occupancy'' in
strongly correlated systems is violated. First, it was pointed out in
Refs.~\cite{Plakida89,Yushankhai91} for the $t$-$J$ model and then in
Ref.~\cite{Plakida14} for the EHM model. This constraint can be  formulated in terms of a
specific  relation for the anomalous (pair) correlation function for the Hubbard
operators. It is easy to verify that a product of two Hubbard operators for the singly
occupied subband equals zero: $\, X_i^{0 \sigma }  X_i^{0 \bar{\sigma}} =
a_{i\sigma}(1- N_{i\bar\sigma})\, a_{i\bar\sigma}(1- N_{i\sigma}) = 0$. Therefore,   the
corresponding single-site pair correlation function should vanish:
\begin{equation}
F_{ii,\sigma} =
  \langle  X_i^{0 \sigma }   X_i^{0 \bar{\sigma}} \rangle =
    \frac{1}{N} \sum_{{\bf q}}\, \langle  X_{\bf q}^{0 \sigma }
    X_{-{\bf q}}^{0 \bar{\sigma}} \rangle  \equiv 0.
\label{1}
\end{equation}
The symmetry of the Fourier-component of the  pair correlation function
$\,F_{\sigma}({\bf q}) = \langle  X_{\bf q}^{0 \sigma } X_{-{\bf q}}^{0 \bar{\sigma}}
\rangle  \,$ has the symmetry of the superconducting order parameter, i.e., the gap
function. For the tetragonal lattice for the $d$-wave pairing $ \,F_{\sigma}(q_x, q_y) =
- F_{\sigma}(q_y, q_x)\,$ and the condition (\ref{1}) after integration over $q_x, q_y$
is fulfilled. For the $s$-wave pairing $ \,F_{\sigma}(q_x, q_y) =  F_{\sigma}(q_y,
q_x)\,$ and the condition (\ref{1}) is violated. The same condition holds for the pair
correlation function for the second Hubbard subband, $\langle  X_i^{\sigma 2}
X_i^{\bar{\sigma}2} \rangle = 0$. Therefore, the $s$-wave pairing in both the Hubbard
subbands is prohibited in the limit of strong correlations.

To overcome the restriction (\ref{1}) in Refs.~\cite{Valkov03}  it was proposed  to
consider the modified time-dependent pair correlation function:
\begin{equation}
\tilde{F}_{ii,\sigma} (t) =
  \int_{- \infty}^{+\infty} d \omega \, {\rm e}^{i\omega t}\,
  \tilde{J}_{ii,\sigma}(\omega) ,
\label{2}
\end{equation}
where
\begin{equation}
  \tilde{J}_{ii,\sigma}(\omega) = {J}_{ii,\sigma}(\omega)-
 \delta(\omega) \;\int_{- \infty}^{+\infty} d \omega_1 \,  J_{ii,\sigma}(\omega_1)  .
\label{3}
\end{equation}
The  spectral density $\,{J}_{ii,\sigma}(\omega)\,$ determines the original correlation
function
\begin{equation}
{F}_{ii,\sigma} (t) =
  \langle  {X_{i}^{0\sigma}}(t)  {X_{i}^{0 \bar{\sigma}}} \rangle =
  \int_{- \infty}^{+\infty} d \omega \, {\rm e}^{i\omega t}\,
 {J}_{ii,\sigma}(\omega) .
\label{4}
\end{equation}
For the modified spectral density  (\ref{3}) the condition  (\ref{1}) is trivially
satisfied for any spectral function $\, J_{ii,\sigma}(\omega) \,$ and the restriction on
the $s$-wave pairing seems to be  lifted. However, thy spectral density (\ref{3}) results
in the nonergodic behavior~\cite{Kubo57}  of the pair correlation function (\ref{2}):
\begin{eqnarray}
C_{ii,\sigma} =  \lim_{t \to \infty} \tilde{F}_{ii,\sigma} (t)
 = - \frac{1}{N} \sum_{{\bf q}}\, \langle X_{\bf q}^{0 \sigma }
    X_{-{\bf q}}^{0 \bar{\sigma}} \rangle ,
\label{5}
\end{eqnarray}
where the conventional pair correlation function decays in the limit ${t \to \infty} $
due to finite life-time effects
\begin{equation}
 \lim_{t \to \infty} {F}_{ii,\sigma} (t)  =
\int_{- \infty}^{+\infty} d \omega \, {\rm e}^{i\omega t}\, J_{ii,\sigma}(\omega) = 0.
\label{6}
\end{equation}
The nonergodic behavior of the modified pair correlation function (\ref{2})  contradicts 
the basic  properties of physical systems and appears for some pathological models with
local integrals of motion~\cite{Suzuki71,Huber77}. In that case, the nonergodic constants
can be found from $1/\omega$ poles of the anticommutator or causal Green functions, as described for the Hubbard model for spin or charge excitations in Refs.~\cite{Mancini03,Avella07} contrary to the arbitrary definition (\ref{3}). Therefore, the statement given in Refs.~\cite{Valkov03}: ``The inclusion of a singular contribution to the spectral intensity of the anomalous correlation function regains the sum rule and remove the unjustified forbidding of the $s$-symmetry order parameter in superconductors with
strong correlations''  cannot be accepted.\\

To conclude, the contradiction between the theoretical and experimental results claimed
in Ref.~\cite{Valkov16}: a more stable  $s$-wave superconducting  pairing with respect to
the intersite Coulomb interaction found in  Refs.~\cite{Zaitsev04, Valkov11}, and a
strong suppression of  the $d$-wave pairing found in Refs.~\cite{Plakida13,Plakida14}, in
fact, is absent. The $s$-wave pairing in the limit of strong correlations is prohibited
due to the kinematical restriction (\ref{1}), while the $d$-wave pairing found within the
EHM can be suppressed only for unphysically large Coulomb interaction $V$, as shown in
Refs.~\cite{Plakida14}. The cancellation of the intersite Coulomb interaction for the
$d$-wave pairing in SFM was found in Ref.~\cite{Valkov16} only  for the nearest-neighbor
oxygen sites, which does not prevent suppression of $d$-wave pairing
due to the  Coulomb interaction for  oxygen sites in different unit cells as in EHM.\\

The author thanks V. Yushankhai for useful discussions.

\end{document}